\documentclass[pra,superscriptaddress,twocolumn,showpacs,preprintnumbers,nofootinbib,amsmath,amssymb,longbibliography]{revtex4-2}

\usepackage{multirow}
\usepackage{graphicx}
\usepackage{dcolumn}
\usepackage{bm}
\usepackage{psfrag}
\usepackage{epsfig}
\usepackage{amsmath}
\usepackage{amssymb}
\usepackage{MnSymbol}
\usepackage{color}
\usepackage{bbm}
\usepackage[FIGTOPCAP,raggedright,nooneline]{subfigure}
\usepackage{verbatim}
 \usepackage[applemac]{inputenc}
\usepackage{textcmds} 
\usepackage{tikz}
\usepackage{natbib}
\usepackage[colorlinks,linkcolor=blue,citecolor=blue,urlcolor=blue,hyperindex,driverfallback=dvipdfm]{hyperref}
\usepackage{ulem}
\usepackage{xcolor}
\usepackage{xspace}

\def\ii{{\rm i}}  \def\ee{{\rm e}}
\def\Ree{{\rm Re}}  \def\Imm{{\rm Im}}
\newcommand{\abs}[1] {\mathopen{}\left|#1\right|\mathclose{}}

\newcommand{\ccpar}[1] {\mathopen{}\left(#1\right)\mathclose{}}
\newcommand{\sqpar}[1] {\mathopen{}\left[#1\right]\mathclose{}}
\newcommand{\clpar}[1] {\mathopen{}\left\{#1\right\}\mathclose{}}
\newcommand{\av}[1]{\left\langle #1 \right\rangle}
\newcommand{\pd}[2] {\mathopen{}\frac{\partial#1}{\partial#2}\mathclose{}}

\def\rb{{\bf r}}  \def\Rb{{\bf R}}  \def\ub{{\bf u}}  
\def\xx{\hat{\bf x}}  \def\yy{\hat{\bf y}}

\def\Eb{{\bf E}}        
      
  \def\jb{{\bf j}}
  
\def\vF{v_{\rm F}}    \def\EF{{E_{\rm F}}}
\def\wp{{\omega_{\rm p}}}  \def\kp{k_{\rm p}}  
    
\def\ww{\omega}
\def\eps{\epsilon}  
\def\epsa{\epsilon_{\rm a}}  \def\epsb{\epsilon_{\rm b}}  
\def\epsr{\epsilon_{\rm r}}  \def\epseff{\epsilon^{\rm eff}}
\def\Hm{\mathcal{H}}
\def\Mm{\mathcal{M}}  \def\Vm{\mathcal{V}}  \def\Dm{\mathcal{D}}  \def\Lm{\mathcal{L}}
    
\def\ub{{\bf u}}
  
\def\vth{{\vec{\theta}}}
      
\def\tht{{\tilde{\theta}}}  
\def\vg{v_{\rm g}}
\def\gamD{\gamma_{\rm D}}

\begin{document}

\title{Nonlinear quantum logic with colliding graphene plasmons}

\author{Giuseppe Calaj\'o}
\email{giuseppe.calajo@pd.infn.it}
\affiliation{ICFO-Institut de Ciencies Fotoniques, The Barcelona Institute of Science and Technology, 08860 Castelldefels (Barcelona), Spain}
\affiliation{Istituto Nazionale di Fisica Nucleare (INFN), Sezione di Padova, I-35131 Padova, Italy.}

\author{Philipp K. Jenke}
\affiliation{University of Vienna, Faculty of Physics, Vienna Center for Quantum Science and Technology (VCQ), Boltzmanngasse 5, 1090 Vienna, Austria}
\affiliation{University of Vienna, Vienna Doctoral School in Physics, Boltzmanngasse 5, 1090 Vienna, Austria}

\author{Lee A. Rozema}
\affiliation{University of Vienna, Faculty of Physics, Vienna Center for Quantum Science and Technology (VCQ), Boltzmanngasse 5, 1090 Vienna, Austria}

\author{Philip Walther}
\affiliation{University of Vienna, Faculty of Physics, Vienna Center for Quantum Science and Technology (VCQ), Boltzmanngasse 5, 1090 Vienna, Austria}
\affiliation{University of Vienna, Research Platform for Testing the Quantum and Gravity Interface (TURIS), Boltzmanngasse 5, 1090 Vienna, Austria}
\affiliation{Christian Doppler Laboratory for Photonic Quantum Computer, Faculty of Physics, University of Vienna, 1090 Vienna, Austria}

\author{Darrick E. Chang}
\affiliation{ICFO-Institut de Ciencies Fotoniques, The Barcelona Institute of Science and Technology, 08860 Castelldefels (Barcelona), Spain}
\affiliation{ICREA-Instituci\'o Catalana de Recerca i Estudis Avan\c{c}ats, 08010 Barcelona, Spain}

\author{Joel~D.~Cox}
\email{cox@mci.sdu.dk}
\affiliation{POLIMA---Center for Polariton-driven Light--Matter Interactions, University of Southern Denmark, Campusvej 55, DK-5230 Odense M, Denmark}
\affiliation{Danish Institute for Advanced Study, University of Southern Denmark, Campusvej 55, DK-5230 Odense M, Denmark}

\date{\today}

\begin{abstract}
Graphene has emerged as a promising platform to bring nonlinear quantum optics to the nanoscale, where a large intrinsic optical nonlinearity enables long-lived and actively tunable plasmon polaritons to strongly interact. Here we theoretically study the collision between two counter-propagating plasmons in a graphene nanoribbon, where transversal subwavelength confinement endows propagating plasmons with 
a flat band dispersion that enhances their interaction. This scenario presents interesting possibilities towards the implementation of multi-mode polaritonic gates that circumvent limitations imposed by the Shapiro no-go theorem for photonic gates in nonlinear optical fibers. As a paradigmatic example we demonstrate the feasibility of a high fidelity conditional $\pi$ phase shift (CZ), where the gate performance is fundamentally limited only by the single-plasmon lifetime. These results open new exciting avenues towards quantum information and many-body applications with strongly-interacting polaritons.
\end{abstract}
 
\maketitle

The integration of a photonic gate within an optical waveguide constitutes a long-standing challenge in quantum optics that is---as elucidated by the Shapiro theorem---tantamount to overcoming practical limitations associated with the large entanglement spread in momentum space produced by multi-photon scattering \cite{shapiro2006single,gea2010impossibility,he2012continuous,dove2014phase}.
Because these restrictions usually apply to co-propagating particles governed by a linear dispersion relation and interacting via a local Kerr-like nonlinearity, they are typically circumvented by invoking nonlocal interactions, which specifically have been exploited in the context of electromagnetically-induced transparency (EIT) with Rydberg atoms \cite{friedler2005long,gorshkov2011photon,bienias2016quantum,shahmoon2011strongly}, as well as in engineered discrete networks of cavities \cite{brod2016passive,combes2018two,borregaard2015heralded,heuck2020controlled,heuck2020photon}
and chiral waveguides \cite{ralph2015photon}.
However, similar strategies have yet to be identified in an integrated photonic platform that can bring quantum optical logic to the nanoscale. 

Plasmon polaritons---quasiparticles that emerge when light hybridizes with collective oscillations of conduction electrons at a metal-dielectric interface---exhibit enhanced dispersion, extending well beyond the light line; the large wave vectors attained by plasmons correspond to an intense concentration of electromagnetic energy on length scales far below the wavelength of the light that excites them \cite{gramotnev2010plasmonics,gramotnev2014nanofocusing,rivera2016shrinking}. Metal nanostructures supporting plasmon resonances have thus been actively explored to enhance nonlinear light-matter interactions on nanometer length scales \cite{kauranen2012nonlinear,tame2013quantum,de2016harmonics}. 
However, both low intrinsic optical nonlinearity and high Ohmic losses encountered in conventional plasmonic materials limit practical implementation of single-photon-level nonlinearity \cite{west2010searching}.

Graphene has recently emerged as a promising material platform for both plasmonics and nonlinear optics: the long-lived \cite{hanson2008dyadic,jablan2009plasmonics,woessner2015highly,ni2018fundamental} and electrically-tunable plasmons supported by the atomically-thin carbon layer can intensify optical near-fields that drive its relatively large intrinsic optical nonlinearity \cite{koppens2011graphene,cox2014electrically,manzoni2015second,cox2019nonlinear}, where the latter attribute stems from a linear electronic dispersion relation that renders charge carrier motion anharmonic \cite{hendry2010coherent,mikhailov2016quantum}. The excitement surrounding the appealing nonlinear and optoelectronic properties of graphene has naturally stimulated efforts to trigger quantum nonlinear optical processes based on plasmon polaritons in integrated nanophotonic platforms \cite{koppens2011graphene,gullans2013single,jablan2015multiplasmon,manzoni2015second}, including plasmon gates \cite{calafell2019quantum} and entangled plasmon pair generation via spontaneous parametric down-conversion \cite{sun2022graphene}.

\begin{figure*}[!t]
    \includegraphics[width=0.9\textwidth]{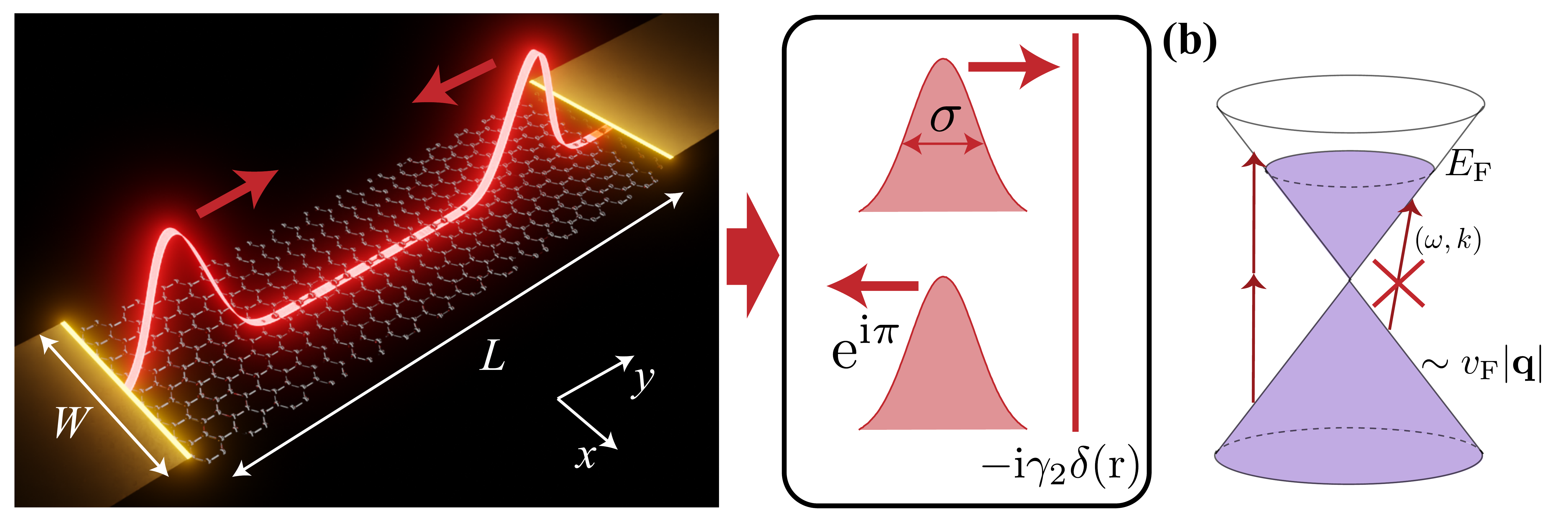}%
    \caption{\textit{Platform}. (a) Two counter-propagating single-plasmon pulses strongly interact in a graphene nanoribbon via a two-plasmon absorption process to acquire a relative $\pi$ phase after the scattering event. The process is mapped in a relative coordinate frame to the simple problem of a massive particle scattered by a delta potential. (b) Sketch of the electronic band structure for doped graphene; when $\hbar\omega<2\EF-\hbar\vF k$, single-plasmon absorption via electron-hole pair excitation is suppressed, while two-plasmon absorption can occur via a nonlinear interband transition. 
    }
    \label{fig:Setup}
\end{figure*}

Here we propose to exploit the strong optical nonlinearity, enhanced by the flat dispersion of guided plasmons in graphene nanoribbons, to collide counter-propagating polaritons and effectively implement an integrated CZ gate within a plasmonic waveguide. 
Our proposal relies crucially on the relatively large propensity for two-plasmon absorption in graphene \cite{jablan2015multiplasmon,mikhailov2016quantum,calafell2019quantum}, which we show here to manifest in the interaction of propagating polaritons as a reflective potential. Such a scenario, in a more generic context, is closely related to a Tonks-Girardeau gas \cite{tonks1936complete,girardeau1960relationship,paredes2004tonks,kinoshita2004observation}, which represents the strongly-interacting limit of the well-known Lieb and Liniger model for massive bosons with contact interaction \cite{lieb1963exact}. 
The CZ gate comprised of colliding plasmons in a graphene nanoribbon is practically limited by the intrinsic single-plasmon absorption rate that is commonly quantified by the quality factor of the associated optical resonance; as we show here, the proposed CZ gate is robust under realistic values commensurate with experiments in graphene plasmonics. 

\section{Model}

We envision a system where incoming single-photon pulses of frequency $\omega$, propagating in a low-loss photonic waveguide, are converted to plasmons via injection into a nonlinear (active) region comprised of a graphene nanoribbon in the $\Rb=(x,y)$ plane with length $L$ and width $W\ll L$, as depicted in Fig.~\ref{fig:Setup}(a). Assuming $W$ to be much smaller than the incident light wavelength, we describe highly-confined plasmons in the quasistatic limit using a scalar potential $\phi_k(x)\ee^{\ii k y}$, which is decomposed in the longitudinal wavevector $k$ to exploit translational invariance in the $y$-direction. Then, defining the graphene nanoribbon by a two-dimensional conductivity $\sigma^{(1)}(\Rb,\ww)=f_\Rb\sigma_\ww^{(1)}$, where $\sigma_\ww^{(1)}$ is the local linear conductivity of extended graphene and $f_\Rb$ is a geometrical parameter that is 1 within the ribbon structure and zero everywhere else \cite{jga2013multiple,christensen2015kerr}, we express the self-consistent potential as $ \phi_k(\theta) = \eta_\ww^{(1)}\mathcal{M}_{kW}\phi_k(\theta)$,
where $\eta_\ww^{(1)}= \ii\sigma_\ww^{(1)}/\ww W$ contains the dependence on size and intrinsic conductivity, $\theta\equiv x/W$ is a normalized coordinate, and
\begin{align}
    \mathcal{M}_q\phi(\theta) \equiv 2 \int &d\theta' K_0(q|\theta-\theta'|) \\
    &\times \clpar{\partial_{\theta'}\sqpar{f_{\theta'}\partial_{\theta'}\phi(\theta')} - q^2f_{\theta'}\phi(\theta')}  \nonumber 
\end{align}
is an integrodifferential operator given in terms of the modified Bessel function $K_0$ and normalized wavevector $q\equiv kW$. Following the prescription of Ref.\ \cite{christensen2015kerr}, we express $\mathcal{M}$ in a discretized real-space basis to extract the eigenvalues $\eta_{n,k}$ and eigenvectors $\phi_{n,k}(\theta)$ that define polaritonic modes supported by the nanoribbon geometry.

\begin{figure}[!t]
    \includegraphics[width=0.9\columnwidth]{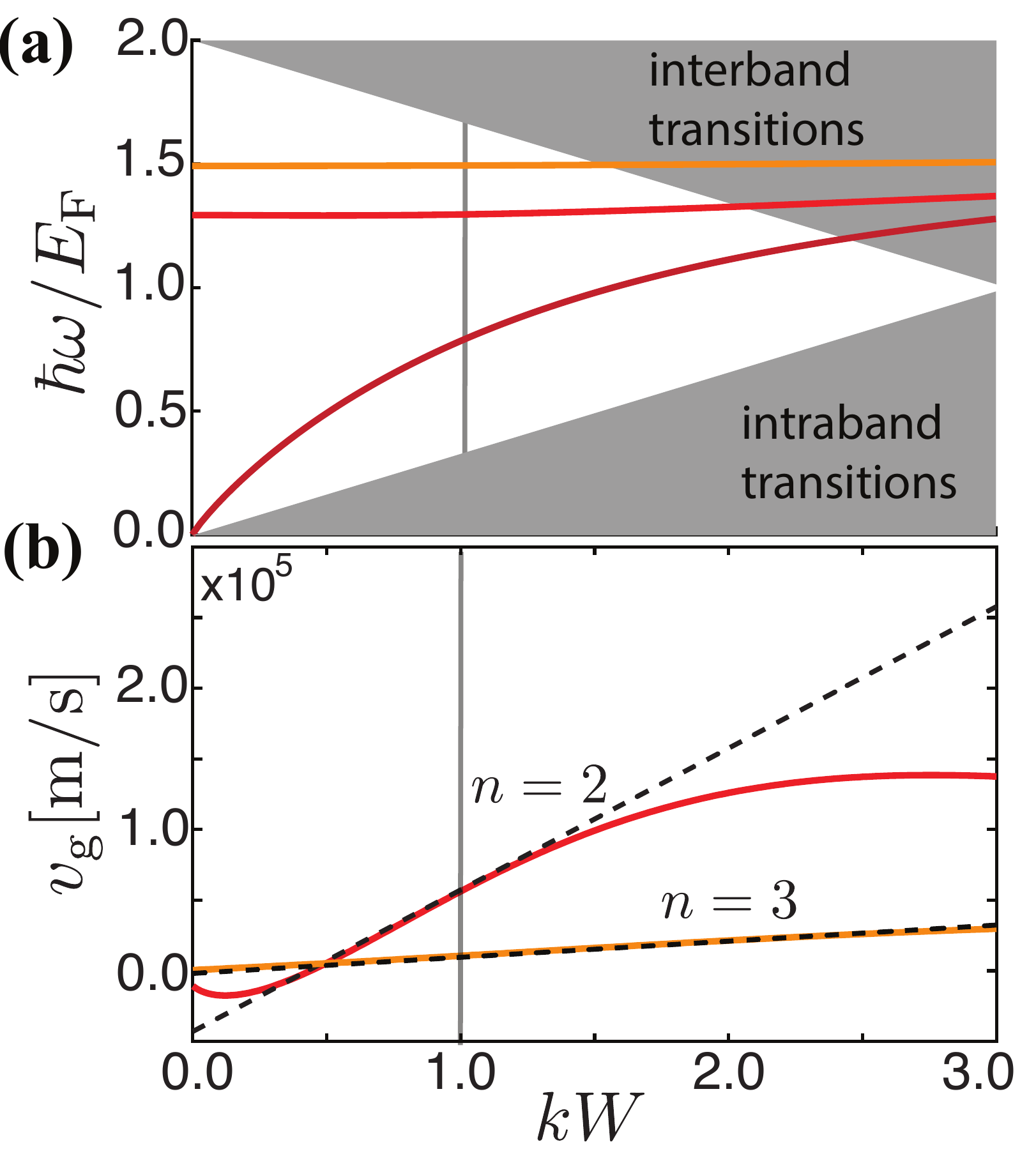}%
    \caption{\textit{Platform}. (a) Plasmon dispersion for the first three modes of a $W=20$\,nm wide nanoribbon doped to a Fermi energy $\EF=0.1$\,eV. The grey areas indicate regions where damping from single-plasmon absorption occurs due to electron-hole pair excitation. (b) Computed plasmon group velocity (solid curves)  compared with that corresponding to a quadratic dispersion, $\vg(k)\simeq \vg(\kp)+\hbar(k-\kp)/m$ (dashed curves). The results are obtained using the linear graphene conductivity obtained in the local random phase approximation, while the vertical line in both panels indicates the plasmon wavevector $kW=1$ considered throughout the proposal. 
   }
    \label{fig:dispersion_main}
\end{figure}

Transverse confinement provided by the ribbon along the $x$-direction leads to distinct branches in the polariton dispersion relation, which we explicitly compute by invoking the linear conductivity of graphene described in the local limit of the random phase approximation (LRPA) \cite{koppens2011graphene},
\begin{equation} \label{eq:LRPA}
    \sigma_\omega^{(1)} = \frac{\ii e^2}{\pi\hbar^2}\frac{\EF}{\ww+\ii\gamD} + \frac{e^2}{4\hbar}\sqpar{\Theta(\hbar\ww-2\EF)+\frac{\ii}{\pi}\log\abs{\frac{\hbar\ww-2\EF}{\hbar\ww+2\EF}}} ,
\end{equation}
where $e$ is the elementary charge and $\EF$ the Fermi energy. The first term in the conductivity accounts for intraband motion of free charge carriers offset by a phenomenological damping rate $\gamD$, while the second term describes interband transitions between the Dirac cones shown in Fig.\ \ref{fig:Setup}(b).
The resulting dispersion for the first three modes is shown in Fig.\ \ref{fig:dispersion_main}(a).
Importantly, the strong light-matter hybridization associated with plasmon polaritons pushes their dispersion well beyond the light line of free-space photons, endowing the propagating quasiparticles with an effective mass $m=\left.\hbar(\partial_k^2\ww_{n,k}\right\vert_{k=\kp})^{-1}$ and a slow group velocity $v_{\rm g}=\left.\partial_k\ww_{n,k}\right\vert_{k=\kp}$ that are characterized by expanding the plasmon dispersion relation around a given plasmon resonance frequency $\wp\equiv\omega_{n,\kp}$ at $k=\pm \kp$ according to $\ww_{n,k}=\wp\pm v_{\rm g}(k\mp\kp)+\hbar/(2m)(k\mp\kp)^2+\dots $.
For small wave vectors, we find that the higher-order polariton modes corresponding to $n>1$ are well-approximated by truncating beyond the quadratic terms in the above expansion, as revealed in Fig.\ \ref{fig:dispersion_main}(b) by the comparison with the numerically-extracted group velocity of the $n=2$ and $n=3$ modes (black dashed curves) for values near $kW\simeq 1$.  

The LRPA conductivity in Eq. \eqref{eq:LRPA} that is used to obtain the dispersion relation predicts a sign change in the imaginary part of the conductivity at the frequency $\omega_{\rm plasma}\simeq 5\EF/3$, which can be interpreted as the plasma frequency beyond which doped graphene ceases to exhibit metallic behavior. Such a feature significantly deviates the plasmon dispersion near and above the Fermi energy $\EF$ from that predicted by the purely-intraband Drude model conductivity (see App.\ \ref{Sec.plasmon dispersion} for details), becoming particularly important for higher-order polariton modes. Here the bands are flattened, leading to large effective polariton masses that can exceed the electron mass of $10^{-31}$\,kg. Incidentally, the engineering of flat bands in graphene nanoribbons presents an alternative to utilizing a Bragg grating for engineering slowly-propagating polariton modes \cite{lu2015graphene}.

Plasmon propagation along the graphene nanoribbon can be described in a second-quantization formalism for massive particles by the effective Hamiltonian 
\begin{equation}\label{eq:H0}
    \Hm_0=-\sum_{\nu={\rm R,L}}\int_0^L d y \hat a_{\nu}^{\dagger}(y)\left(
\frac{\hbar^2}{2m}\frac{\partial^2}{\partial y^2}\pm \ii\hbar \bar v_{\rm g}\frac{\partial}{\partial y}\right)\hat a_{\nu}(y),
\end{equation}
where $\hat a_{\rm R(L)}(y)$ and $\hat a_{\rm R(L)}^{\dagger}(y)$ are the bosonic field operators respectively annihilating and creating a right-propagating (left-propagating) plasmon at position $y$ and $\bar v_{\rm g}=v_{\rm g}-(\hbar/m)\kp$. 
Graphene plasmons are understood as well-defined and long-lived excitations only in the absence of incoherent scattering processes involving phonons or defects, and can be further damped by electron-hole pair excitation channels \cite{jablan2009plasmonics}. Fortunately, owing to the Pauli principle, the latter process is suppressed in highly-doped graphene, where absorption via intraband and interband transitions is prohibited for plasmon energies within $\hbar\vF k<\hbar\wp<2\EF-\hbar\vF k$, as illustrated in Fig.\ \ref{fig:Setup}(b). The remaining decoherence mechanisms, mainly related to phonon and defect scattering \cite{jablan2009plasmonics}, are incorporated in the single-plasmon absorption rate $\gamma_1=\wp/Q$, which we characterize by the quality factor $Q$. The incoherent scattering processes captured in $\gamma_1$ limit plasmon propagation by imposing a decay $\ee^{-2\gamma_1\tau}$, where $\tau=L/v_{\rm g}$ represents the time it takes to propagate over a distance $L$; as will be discussed later, single-plasmon absorption mainly affects the free evolution of plasmons and not their interaction dynamics.  

Beyond single-plasmon absorption, plasmon \textit{pairs} can be efficiently absorbed within a certain frequency range via interband transitions \cite{jablan2015multiplasmon,calafell2019quantum}, as illustrated in Fig.\ \ref{fig:Setup}(b). The associated nonlinear process is encoded in the real part of the third order conductivity $\sigma_\omega^{(3)}$, which admits analytical expression in the local limit of the random phase approximation \cite{mikhailov2016quantum}. In a one-dimensional ribbon, this process is effectively captured by a per-length two-plasmon absorption (TPA) rate that we estimate as
\begin{equation}\label{eq:TPA}
    \gamma_2 = \frac{\hbar\omega_{\rm p}^3\Ree\clpar{\sigma_{\omega_{\rm p}}^{(3)}} \xi^{(3)}_{\kp}}{\left[\Imm\left.\clpar{\sigma_\omega^{(1)}-\ww\partial_\ww\sigma_\omega^{(1)}}\right\vert_{\ww=\ww_p}\xi^{(1)}_{\kp}\right]^2}
\end{equation}
where $\xi^{(j)}_{\kp}= \int dx \abs{\ub_{\kp}(x)}^{j+1}$ is a factor depending on the integral of the electric field mode functions $\ub_{\kp}(x)$ (see App.\ \ref{App A} for further details). Crucially, the rate of TPA in graphene can exceed the single-plasmon decay rate, leading to strong nonlinear effects \cite{jablan2015multiplasmon,calafell2019quantum}.
We model the nonlinear absorption-induced interaction as a dissipative local Kerr nonlinearity described by the effective Hamiltonian for propagating massive interacting plasmons
\begin{equation}\label{eq:H}
    \Hm=\Hm_0-\ii\frac{\gamma_2}{2}\sum_{\nu\nu'={\rm R,L}}\int dy \hat a_{\nu}^{\dagger}(y)\hat a_{\nu'}^{\dagger}(y)\hat a_{\nu}(y)\hat a_{\nu'}(y).
\end{equation}
Note that although the Hamiltonian of Eq.\ \eqref{eq:H} is formally non-Hermitian, it entirely captures (in the absence of external energy sources) the dynamics occurring within a given excitation subspace.

\begin{figure*}[!t]
\includegraphics[width=1.0\textwidth]{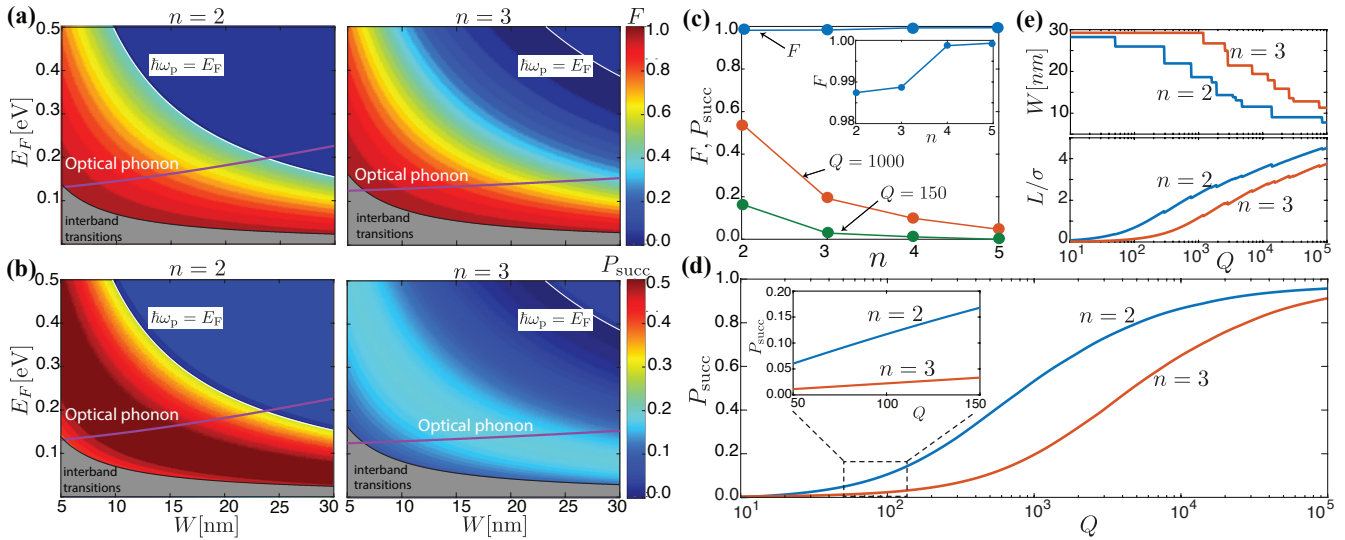}%
\caption{\textit{Gate performance}. (a) Fidelity and (b) success probability as functions of the nanoribbon width $W$ and Fermi energy $\EF$ for the second ($n=2$, left panel) and third ($n=3$, right panel) plasmon modes. White curves indicate the plasmon energy $\hbar\omega=\EF$, while the purple lines show the optical phonon energy in graphene; grey areas indicates regions where interband transitions become important and the plasmon is no longer a long-lived excitation. In each plot the ribbon length was selected to yield a higher success probability, and in (b) the quality factor was set to $Q=1000$. (c) Fidelity and success probability (with $Q=150$ and $Q=1000$) for different dispersion modes ($n$) obtained by optimizing over the parameter range of $\EF$ and $W$ used in panels (a) and (b); the inset shows a zoom of the Fidelity. (d) Optimal success probability as function of $Q$ for the second and third modes; the inset shows a zoom of the area with $Q\in[50,150]$ indicated by the black dashed box. To obtain the curves we fixed $\EF=0.1$\,eV and we optimized with respect to the ribbon width and length. The optimal ribbon width $W$ (in nanometers) and length $L$ (normalized respect to the pulse width) used to obtain the curves in (d) are shown in panel (e). In all plots the pulse wavevector was set to $kW=1$ and the width to $\sigma=W/\Delta k$ with $\Delta k=0.9$.}
\label{fig_F}
\end{figure*}

\section{Two-plasmon dynamics}
Local Kerr nonlinearities are known to lead to a no-go theorem for the implementation of gates among co-propagating photons in nonlinear optical fibers \cite{shapiro2006single,gea2010impossibility,dove2014phase,combes2018two}.
To circumvent this limitation, we consider the peculiar case of a strong hard core collision between two slow counter-propagating plasmons interacting in the active region via TPA, as depicted in Fig.\ \ref{fig:Setup}(a).
Importantly, we specifically consider the higher ($n>1$) dispersion branches, where the flat plasmon dispersion diminishes the contribution of the kinetic term in the Hamiltonian, thus enhancing interactions among plasmons.
The collision between two plasmons can be conveniently described in a relative coordinate frame that effectively maps the interaction to the simple problem of a single massive particle scattered by a delta function potential, as sketched in Fig.\ref{fig:Setup}(a) (see App.\ \ref{App.scattering} for details).
The single-particle scattering is then described by the scattering matrix $S=\int dk (r|-k\rangle\langle k|+t|k\rangle\langle k|)$ that yields the output state $|\psi_{\rm out}\rangle=S |\psi_{\rm in}\rangle$ corresponding to an incoming input state $|\psi_{\rm in}\rangle$, where
\begin{subequations} \label{Eq:r_t}
\begin{align}
    r &= -\frac{1}{1+2\pi\lambda_{\rm a}/\lambda_{\rm p}} , \\
    t &= \frac{1}{1+\lambda_{\rm p}/(2\pi\lambda_{\rm a})} ,
\end{align}
\end{subequations}
are the reflection and the transmission coefficients that depend on the ratio between the plasmon wavelength $\lambda_{\rm p}=2\pi/|\kp|$ and the \textit{absorption length}
\begin{equation}\label{Eq:abs_length}
    \lambda_{\rm a}=\frac{2}{\gamma_2}\left(\frac{2\vg}{|\kp|}-\frac{\hbar}{m}\right)
\end{equation}
associated with a two-plasmon absorption-induced interaction. Incidentally, Eq.\ \eqref{Eq:abs_length} explicitly shows that the band flattening enhances plasmon interactions, with perfect refection (and thus perfect repulsive collision in the original frame) occurring when $\lambda_{\rm a}=0$ is satisfied by the group velocity $\vg=\hbar \kp/(2m)$ that corresponds to a completely flat dispersion for the considered terms up to the second order in the original dispersion expansion.
Importantly, a reflection from the potential in the single-particle picture is accompanied by a $\pi$-phase shift (minus sign in the reflection coefficient of Eqs.\ \eqref{Eq:r_t}), which corresponds to a relative $\pi$-phase shift between the two colliding plasmons.
Note that the transmission and reflection probabilities $T=|t|^2$ and $R=|r|^2$ do not perfectly sum to one since the two-plasmon interaction is inherently dissipative, and lead to a large two plasmon absorption at intermediate $\lambda/\lambda_{\rm a}$ ratios.
%
%

\section{CZ plasmon gate}

As mentioned in the previous section, the key idea underlying our gate proposal is based on the observation that the quasi-flat bands of the higher ($n>1$) modes can strongly enhance the plasmon interaction, leading, in the \qq{Zeno}-like limit of $\lambda_{\rm p}/\lambda_{\rm a}\gg 1$ \cite{franson2004quantum,facchi2004unification}, to an almost perfect collision that endows the plasmons with a relative (conditional) $\pi$-phase shift and thus implementing a CZ gate.
An initial figure of merit to estimate the efficiency of this process is given by the reflection probability, $R=|r|^2$, which for a ratio of $\lambda_{\rm p}/\lambda_{\rm a}\sim 10^{3}$ is on the order of $R\sim 0.99$. A more realistic estimate for gate performance can be achieved by considering two counter-propagating Gaussian pulses, $|\psi_{\rm in}\rangle=\int dk\psi(k)|k\rangle$ with $\psi(k)=\sqrt{\sigma/\sqrt{\pi}}\ee^{-(k-k_0)^2\sigma^2/2}$, where $k_0=\kp$ is the central wavevector and $\sigma$ the pulse width. A quantitative figure of merit for the efficiency of the gate is then given by the ideal state fidelity $F=|\langle\psi_{\rm in}(-k_0)|\psi_{\rm out}(k_0)\rangle|^2$, which is merely the probability to obtain the same initial pulse reflected up to a phase shift, and can be straightforwardly computed by calculating the output state from the $S$-matrix. The values of the computed fidelity as function of the nanoribbon width and Fermi energy are shown in Fig.\ \ref{fig_F}(a), which exhibits isolines with similar fidelities corresponding to the same plasmon frequency, e.g., the isoline associated with $\hbar\wp=\EF$ is shown by the white curve in the figure.
%


The computed fidelity evaluates the efficiency of the scattering process without taking into account single-plasmon absorption during propagation of the two pulses. As discussed above, the main sources of single-plasmon absorption are interband electron-hole pair excitation and phonon scattering; the former process is highly suppressed if the plasmon frequencies lie below the  interband transition region (above the grey area in Fig.\ \ref{fig_F}), while the latter effect is strongly reduced when the plasmon frequency lies below the optical phonon line at $\hbar\omega_{\rm ph}<0.2$\,eV (below the purple line in Fig.\ \ref{fig_F}) \cite{jablan2009plasmonics}. Importantly, Fig.\ \ref{fig_F} shows that high fidelities of $F>0.98$ can be achieved for plasmon frequencies away from both regions. 
The average plasmon lifetime is then given by the single-plasmon decay rate $\gamma_1=\wp/Q$ in terms of a quality factor $Q$ that accounts for all remaining sources of damping, mainly attributable to defect scattering \cite{jablan2009plasmonics}. In this regime, large quality factors $Q>100$ are within experimental capability, while even higher values up to $Q\sim 1000$ have been theoretically predicted \cite{ni2018fundamental}.

Having encoded the remaining sources of damping into the quality factor $Q$, we can define a more accurate estimation of the gate performance in terms of the success probability $P_{\rm p}$.
To estimate this quantity we first observe that, as previously mentioned, the signal detection probability is overall damped  by an exponential factor $\ee^{-2\gamma_1\tau}$, where $\tau=L/\vg$ is the time required for a plasmon to traverse the ribbon. This contribution clearly enhances the probability of plasmon absorption during the gate protocol and penalizes \qq{too slow} group velocities, such as those associated with higher-order plasmon dispersion branches.
While such losses are clearly reduced in shorter ribbons, any benefit must be compared against the spatial length ${\sim}\sigma$ of the counter-propagating pulses; when $\sigma \gtrsim L$, there is a significant probability that the two excitations never overlap in the ribbon at the same time. On the other hand, one could reduce $\sigma$, but this increases the frequency bandwidth, while the large phase shift only occurs within a limited bandwidth.
We account for this trade off by estimating the probability of having a single plasmon within the length $L$, which for a Gaussian pulse reads: $P_{\rm p}=\int^{L/2}_{-L/2} d y|\psi(y-\Delta L)|^2=\left[{\rm{erf}}\left(\frac{\Delta L+L/2}{2\sigma}\right)-\left(\frac{\Delta L-L/2}{2\sigma}\right)\right]$, where we included the effect of a relative delay of the two pulses assuming that  the scattering event occurs at a distance $\Delta L$ from the middle of the ribbon. 
In this way we define the success probability 
\begin{equation}
    P_{\rm succ}=F\ee^{-2\gamma_1\tau}\frac{1}{2}\left[{\rm erf}\left(\frac{\Delta L+L/2}{2\sigma}\right)-\left(\frac{\Delta L-L/2}{2\sigma}\right)\right] ,
\end{equation}
which is shown in Fig.\ \ref{fig_F}(b) for the second and third mode using an optimal ribbon length, assuming a collision at the center of the ribbon, $\Delta L=0$, and fixing the quality factor to $Q=1000$. Note that for the ribbon lengths considered, $L\sim \sigma$, acquired delays of the order of $\Delta L/L=0.1$  affect the predicted success probability by less than $1\%$ with respect to a perfect central collision. Thus we did not explicitly take into account these deviations in Fig.\ \ref{fig_F}(b).

The optimal success probability achievable in each mode (within the low loss region) is plotted in Fig.\ \ref{fig_F}(c) for two different quality factors, and exhibits an opposite trend with respect to the one exhibited in the case of optimal fidelity (blue dots), i.e., becoming more damped for slower, more massive plasmons. Such behavior suggests that even if extremely high fidelities $F\simeq 0.99$ can be ideally reached by higher-order modes, the second and the third modes are the ones exhibiting optimal operational conditions. Indeed, a convenient trade off between high conditional fidelities and good gate success probability can be achieved for quality factors of $Q_f=150$ and $Q_f=1000$, which we predict to be respectively on the order of $P_{\rm succ}\sim 20\%$ and $P_{\rm succ}\sim 50\%$ for the second mode. A broader panoramic over the possible success probabilities achievable with the two considered modes for different quality factors is shown in Fig.\ \ref{fig_F}(d), while the corresponding optimal ribbon width and length are plotted in Fig.\ \ref{fig_F}(e).
Fig.\ \ref{fig_F}(d) illustrates how the gate performance progressively approaches the ideal lossless one for large quality factors, $Q\gtrsim 10^4$. Importantly, in the range of quality factors $Q\in[50,150]$ achievable in realistic devices shown in the inset of Fig.\ \ref{fig_F}(d), success probabilities ranging between $P_{\rm succ}\sim 5-20\%$ can be reached. This makes our proposal comparable with state-of-the-art platforms for implementing photon-photon gates, e.g. cavity QED \cite{hacker2016photon}.
These results indicate that overall good gate performance can be obtained in the proposed setup, with single-plasmon absorption representing the primary limitation.

 Our discussion up to now does not account for possible 
in-plane scattering processes induced by defects or disorder in the structure \cite{garcia2013scattering}, which may eventually lead to Anderson localization \cite{anderson1958absence}. 
For massive particles, the Anderson localization length roughly scales as $L_{\rm loc}^{-1}\sim W[\epsilon/v_{\rm g}(k_p)]^2$ \cite{billy2008direct,sanchez2007anderson,thouless1973localization}, where $\epsilon$ characterizes the disorder strength. This result shows that the effects of disorder are enhanced by slow-light effects. While we do not know of any straightforward way to determine $\epsilon$ for a realistic graphene system based on simple physical considerations, we note that the localization length is in principle an experimentally mesurable quantity. For our purposes, it is then clearly important to ensure that the sample is clean enough to have an associated localization length larger than the ribbon active region, i.e., $L_{\rm loc}\gtrsim L$.


\section{General considerations and conclusions}

In summary, we have shown that by engineering graphene nanoribbons it is possible to induce flat dispersion that enhances the plasmon interaction originating from nonlinear absorption. In the limit of strong interactions, the model effectively behaves as a Tonks-Girardeau gas \cite{tonks1936complete,girardeau1960relationship}, such that counter-propagating plasmons undergo elastic collisions that can be exploited to implement an integrated CZ gate. To describe the plasmon collision process, we have introduced a phenomenological Hamiltonian in Eq.\ \eqref{eq:H} that strongly simplifies an otherwise complex many body-problem. In particular, we treat 2D plasmons---collective charge oscillations dressed by the electromagnetic field---as well-defined bosonic excitations interacting locally via TPA at a fixed per-length rate $\gamma_2$. The present simplifying assumption is informed by current knowledge of nonlinear optical processes in graphene, and should be reasonable within a small frequency range, such as that considered in our proposal involving flat plasmon dispersion branches around $\wp\simeq 1.5 \EF$. The assumption of local interactions can 
be further validated by considering that the natural length of the plasmon interaction should be set by the Fermi length $L_{\rm F}=2\pi \vF/\EF$, which, for Fermi energies of $\EF\simeq 0.1$, is on the order of $L_{\rm F}\simeq 40$\,nm and thus commensurate with the considered pulse widths.

Even if theoretically feasible, the current proposal presents important experimental challenges, which are not only restricted by the fabrication of high-quality nanostructures. Indeed, the precise excitation of individual propagating plasmons is still at the edge of the state-of-the art; nevertheless, promising proposals exist to efficiently couple far-field light to propagating plasmons \cite{nikitin2014efficient}, while experiments have demonstrated almost perfect absorption into acoustic graphene plasmons. On the other hand, the occurrence of strong nonlinear multi-plasmon absorption \cite{jablan2015multiplasmon} could make the current proposal realizable with weak coherent plasmon pulses. We also remark that, for moderately doped ribbons of width $\gtrsim10$\,nm considered in our proposal, the gate performance will be only minimally affected by edge-termination geometry (i.e., armchair or zigzag edges) \cite{cox2016quantum}, although ribbons with armchair edges are preferable to avoid additional plasmon damping due to the presence of edge states in the electronic spectrum of zigzag ribbons.
We finally note that the model described by \eqref{eq:TPA} constitutes a rather generic description of massive particles interacting via a contact-like interaction in one dimension, with the fundamental ingredient relying on the strong ratio between the interaction and the kinetic term. Such ideas could be then implementable also in other suitable platforms such as nonlinear resonator arrays in circuit QED \cite{wang2020supercorrelated}, atomic waveguides \cite{masson2020atomic}, newly-available 2D material heterostructures \cite{low2017polaritons,tartakovskii2020excitons}, and confined excitons in TMDs \cite{thureja2022electrically}.\\

\section{Acknowledgments}
The authors thank Gian Marcello Andolina for valuable discussions. G.~C. acknowledges that results incorporated in this standard have received funding from the European Union Horizon 2020 research and innovation programme under the Marie Sklodowska-Curie grant agreement No 882536 for the project QUANLUX and from the T-NiSQ consortium agreement financed by QUANTERA 2021.
P.~K.~J., L.~A.~R., and P.~W. acknowledge support from the European Union's Horizon 2020 research and innovation programme under grant agreement No. 820474 (UNIQORN) and No 899368) (EPIQUS), the Marie Sklodowska-Curie grant agreement No. 956071 (AppQInfo), and the QuantERA II Programme under Grant Agreement No. I6002-N (PhoMemtor);
the Austrian Science Fund (FWF) through [F7117] (BeyondC), and [FG5];
the AFOSR via FA8655-20-1-7030 (PhoQuGraph), and FA9550-21- 1-0355 (QTRUST); the Austrian Federal Ministry for Digital and Economic Affairs, the National Foundation for Research, Technology and Development and the Christian Doppler Research Association.
D.~E.~C. acknowledges support from the European Union Horizon 2020 research and innovation programme, under European Research Council grant agreement No 101002107 (NEWSPIN); the Government of Spain (Europa Excelencia program EUR2020-112155, Severo Ochoa program CEX2019-000910-S, and MICINN Plan Nacional Grant PGC2018-096844-B-I00), Generalitat de Catalunya through the CERCA program, AGAUR Project No.~2017-SGR-1334, Fundaci\'o Privada Cellex, Fundaci\'o Mir-Puig, and Secretaria d'Universitats i Recerca del Departament d'Empresa i Coneixement de la Generalitat de Catalunya, co-funded by the European Union Regional Development Fund within the ERDF Operational Program of Catalunya (project QuantumCat, ref. 001-P-001644).
J.~D.~C. is a Sapere Aude research leader supported by VILLUM FONDEN (grant no. 16498) and Independent Research Fund Denmark (grant no. 0165-00051B).
The Center for Polariton-driven Light--Matter Interactions (POLIMA) is funded by the Danish National Research Foundation (Project No.~DNRF165).

\section*{Data availability} The data produced in this manuscript can be provided upon request.

\appendix

\section{Optical response of structured 2D materials in the quasistatic limit}\label{App A}

Invoking the quasistatic approximation, we describe the optical properties of a nanostructured two-dimensional (2D) material in the $\Rb=(x,y)$ plane using the scalar potential
\begin{equation} \label{eq:Phi}
    \Phi(\rb,\ww) = \frac{1}{\epseff}\int d^2\Rb' \frac{\rho^{\rm ind}(\Rb',\ww)}{\abs{\rb-\Rb'}} ,
\end{equation}
where $\epseff=(\epsa+\epsb)/2$ accounts for screening of the induced charge density $\rho^{\rm ind}$ by dielectric media with permittivity $\epsa$ and $\epsb$ respectively above and below the 2D material. The induced charge is obtained from the continuity equation $\rho^{\rm ind}(\Rb,\ww) = -(\ii/\ww)\nabla_\Rb\cdot\jb(\Rb,\ww)$, while the use of Ohm's law $\jb=\sigma^{(1)}\Eb$ and $\Eb=-\nabla\Phi$ allows us to express the potential in Eq.\ \eqref{eq:Phi} self-consistently as
\begin{equation}
    \Phi(\rb,\ww) = \frac{\ii}{\epseff\ww}\int \frac{d^2\Rb'}{\abs{\rb-\Rb'}}\nabla_{\Rb'}\cdot\sqpar{\sigma^{(1)}(\Rb',\ww)\nabla_{\Rb'}\Phi(\Rb',\ww)} ,
\end{equation}
where $\sigma^{(1)}(\Rb,\ww)$ is an isotropic 2D conductivity characterizing the intrinsic linear optical response of the 2D material. Following the method of Ref.~\cite{yu2017universal}, the 2D nanostructure morphology is contained in the conductivity by assuming it has the form $\sigma^{(1)}(\Rb,\ww)=f_\Rb\sigma_\omega^{(1)}$, where $f_\Rb$ is 1 within the 2D structure but zero everywhere else, and the potential within the 2D material is expressed in terms of a reduced 2D coordinate vector $\vth=\Rb/D$ as
\begin{equation} \label{eq:Phi_th}
    \Phi(\vth,\ww) = \eta_\omega^{(1)} \int d^2\vth' \frac{1}{\abs{\vth-\vth'}}\nabla_{\vth'}\cdot\sqpar{f_{\vth'}\nabla_{\vth'}\Phi(\vth',\ww)} ,
\end{equation}
where the dimensionless parameter $\eta_\omega^{(1)} = \ii\sigma^{(1)}_\ww/\epseff\ww D$ contains all dependence on the intrinsic conductivity of the 2D material (in the local limit), its dielectric environment, and characteristic size $D$ (e.g., the diameter of a disk or the side length of a square).
\subsection{Guided modes in 2D nanoribbons}

We characterize the 2D nanoribbon geometry by a finite width $W$ along $\xx$ but infinite extension in $\yy$, so that translational invariance in the latter dimension suggests a Fourier decomposition of the potential in wave vector components $k$ according to $\Phi(\Rb)=\phi(x)\ee^{\ii k y}$. Following the prescription of Ref.\ \cite{christensen2015kerr}, we write the self-consistent potential of Eq.\ \eqref{eq:Phi_th} in terms of the normalized coordinate $\theta\equiv x/W$ as
\begin{equation} \label{eq:phi_eig}
    \phi(\theta)= \eta_\omega^{(1)} \Mm \phi(\theta)  ,
\end{equation}
where $\Mm=\Vm\Dm$ is a product of the differential operator
\begin{equation}
    \Dm\phi(\theta)\equiv \clpar{\partial_\theta\sqpar{f_\theta\partial_\theta\phi(\theta)}-k^2W^2f_\theta\phi(\theta)}
\end{equation}
and an integral operator
\begin{equation}
 \Vm \phi(\theta)  \equiv =2\ee^{\ii k y}\int d\theta' K_0(kW\abs{\theta-\theta'})\phi(\theta')
\end{equation}
involving the modified Bessel function $K_0$. Discretizing $\theta\in[0,1]$ in $N$ equally-spaced points as $\{\theta_l\}_{l=1}^N$ such that $h=\theta_{l+1}-\theta_l$ for all $l$, we assign $\phi_l\equiv\phi(\theta_l)$ and $f_l\equiv f(\theta_l)$ to represent $\Dm$ as $\Dm\phi_l = \sum_{l'}D_{ll'}\phi_{l'}$, where
\begin{equation}
\begin{split}
    D_{ll'} = &\frac{1}{2h^2}\left[\delta_{l-1,l'}\ccpar{f_{l-1}+f_l} - \delta_{ll'}\ccpar{f_{l-1}+2f_l+f_{l+1}} \right.\\
  &  \left.+ \delta_{l+1,l'}\ccpar{f_l+f_{l+1}} - \delta_{ll'}f_lk^2W^2\right]
    \end{split}
\end{equation}
is obtained using central differences \cite{christensen2015kerr}, while the boundary condition associated with the vanishing of normal current $\left.\partial_\theta\phi(\theta)\right\vert_{\theta=0}=\left.\partial_\theta\phi(\theta)\right\vert_{\theta=1}=0$ at the ends of the ribbon leads to
\begin{subequations}
\begin{align}
    D_{1l'} &= \frac{1}{2h^2}\ccpar{f_1+f_2}\ccpar{-\delta_{1l'}+\delta_{2l'}} - \delta_{1l'} f_1 k^2 W^2  \\
    D_{Nl'} &= \frac{1}{2h^2}\ccpar{f_{N-1}+f_N}\ccpar{\delta_{N-1,l'}-\delta_{Nl'}} - \delta_{Nl'} f_N k^2 W^2 .
\end{align}
\end{subequations}
Meanwhile, assuming a slowly-varying $\phi(\theta)$, the matrix decomposition of $\Vm$ is
\begin{equation} \label{eq:V_llp}
\begin{split}
    V_{ll'} &= 2\int_{\theta_{l'}-h/2}^{\theta_{l'}+h/2} d\theta' K_0\ccpar{q\abs{\theta_l-\theta'}} = \\
    &\sum_{\tht=\theta_{ll'}\pm h/2}(\pm\pi )\tht\sqpar{K_0\ccpar{q\abs{\tht}}\Lm_{-1}\ccpar{q\abs{\tht}} + K_1\ccpar{q\abs{\tht}}\Lm_0\ccpar{q\abs{\tht}}} ,
    \end{split}
\end{equation}
where $\theta_{ll'}\equiv\theta_l-\theta_{l'}$, $q\equiv kW$ is the normalized wave vector, and $\Lm_n$ denotes the modified Struve function of order $n$ \cite{abramowitz1948handbook}. Note that in the $q\to0$ limit, charge neutrality enables replacement of $K_0\ccpar{q\abs{\theta-\theta'}}\to - \log\abs{\theta-\theta'}$ in the kernel of the operator $\Vm$, so that Eq.\ \eqref{eq:V_llp} becomes
\begin{equation}
    V_{ll'} = 2\sum_\pm(\pm)\ccpar{\theta_{ll'}\pm h/2}\ccpar{1 - \log\abs{\theta_{ll'}\pm h/2}} .
\end{equation}
\subsection{Plasmon dispersion in graphene nanoribbons}\label{Sec.plasmon dispersion}

\begin{figure}[!t]
\includegraphics[width=0.8\columnwidth]{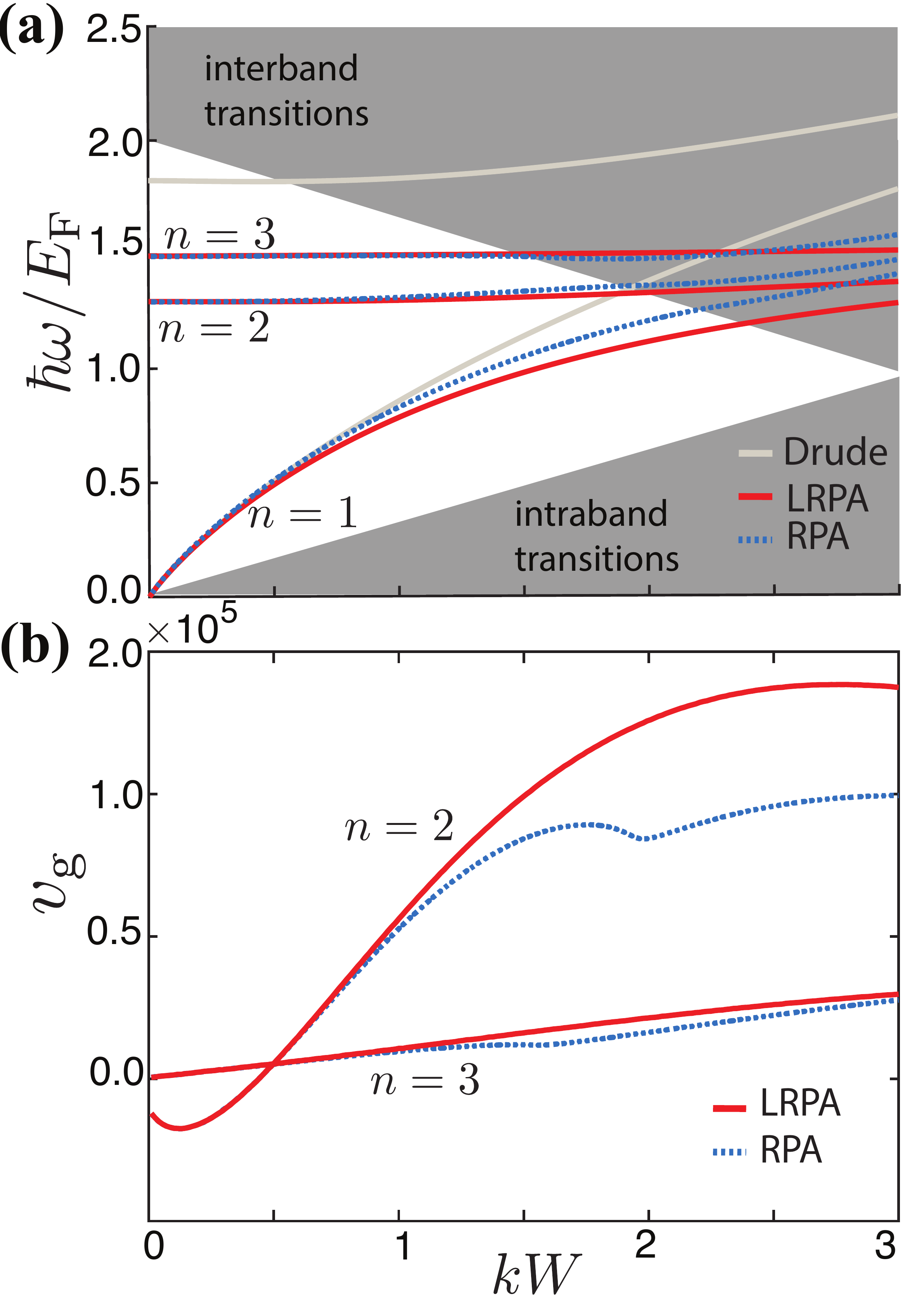}%
\caption{(a) Dispersion relation for the first three plasmon modes supported by a graphene nanoribbon of width $W=20$\,nm and doped to a Fermi energy $\EF=0.1$\,eV as predicted using the Drude model (grey curves), the local random phase approximation (LRPA, red curves), and the full RPA taking into account nonlocal effects (RPA, blue dots). The black dashed line indicates the threshold where single-plasmon absorption via Landau damping kicks in. (b) Group velocity $\vg$ for the second and third modes in (a) as function of the normalized wavevector in the LRPA and RPA. }
\label{fig:dispersion}
\end{figure}

The solution of $\Mm\phi_{n,k}=\eta_{n,k}^{-1}\phi_{n,k}$ yields the eigenmodes $\phi_{n,k}(\theta)$ and eigenvalues $\eta_{n,k}^{-1}$ of the nanoribbon geometry, while comparison to Eq.\ \eqref{eq:phi_eig} leads to the dispersion condition
\begin{equation} \label{eq:dispersion}
 -\eta_{n,k} = \frac{{\rm Im}\{\sigma_{\omega_{n,k}}^{(1)}\}}{\epsilon^{\rm eff}\omega_{n,k} W}.
\end{equation}
In the main text, the dispersion relation of mode $n$ in a graphene nanoribbon is found by inserting the conductivity of extended graphene in Eq.\ \eqref{eq:LRPA}, obtained within the local limit of the random-phase approximation (RPA).
As mentioned in the main text, the presence of the interband term in the LRPA conductivity modifies the dispersion relation respect to that obtained with the Drude model.
For a direct comparison, the solutions of Eq.\ \eqref{eq:dispersion} are plotted as red curves in Fig.\ \ref{fig:dispersion} for the local-RPA (LRPA) conductivity given in Eq.\ \eqref{eq:LRPA}, contrasted with the solutions obtained solely from the intraband contribution (Drude) or by adopting a more sophisticated model that incorporates nonlocal effects (RPA), i.e., for $\sigma_\omega^{(1)}\to\sigma^{(1)}(k,\ww)$ in Eq.\ \eqref{eq:dispersion}, where $\sigma^{(1)}(k,\ww)$ is reported in Ref.\ \cite{koppens2011graphene}.
The zero of the imaginary part of the LRPA conductivity at $\hbar\omega_{\rm plasma}\simeq 5\EF/3$ flattens the plasmon bands compared to the simple Drude case (see Fig.\ \ref{fig:dispersion}(a)), which instead captures the dispersion only in the low-frequency $\hbar\ww<\EF$ regime. On the other hand, nonlocal effects captured in the full RPA conductivity emerge at higher frequencies $\hbar\ww\gtrsim\EF$, becoming particularly prominent for large plasmon wave vectors $\hbar k\vF\sim \EF$; the effect of these corrections is shown in Fig.\ \ref{fig:dispersion}(b), which indicates that the LRPA conductivity faithfully captures the plasmon dispersion for wave vectors $kW\lesssim 1$ considered in this work.

\subsection{Mode normalization}

The modes supported by a graphene nanoribbon are normalized according to well-established procedures for quantizing the electromagnetic field in dispersive media \cite{bhat2006hamiltonian,ferreira2020quantization}, leading to the condition
\begin{equation}
    \frac{1}{4\pi}\Ree\clpar{\int d^3\rb \Eb_\ww^*(\rb) \sqpar{\eps(\rb,\ww) + \frac{\ww}{2}\pd{}{\ww}\eps(\rb,\ww)} \Eb_\ww(\rb)} = \frac{\hbar\ww}{2},
\end{equation}
where
\begin{equation}
    \eps(\rb,\ww) = \epsr + \frac{4\pi\ii}{\ww}\sigma^{(1)}(\Rb,\ww)\delta(z)
\end{equation}
is the dielectric function associated with a single 2D layer characterized by a surface conductivity $\sigma$ embedded in a homogeneous medium with relative dielectric permittivity $\epsr$. Here we again assume a conductivity of the form $\sigma(\Rb,\ww)=f_\Rb\sigma_\omega^{(1)}$, so that the nanoribbon geometry is described by the dimensionless factor $f_\Rb$ that takes a value of unity within the structure and zero everywhere else; the normalization condition is now
\begin{equation}
\begin{split}
    &\frac{1}{4\pi}\Ree\left\{\ccpar{\epsr + \frac{\ww}{2}\pd{\epsr}{\ww}} \int d^3\rb \Eb_\ww^*(\rb)\cdot\Eb_\ww(\rb)\right. +\\ & \left.\frac{2\pi\ii}{\ww}\pd{}{\ww}\ccpar{\ww\sigma_\omega^{(1)}} \int d^2\Rb f_\Rb \Eb_\ww^*(\Rb)\cdot\Eb_\ww(\Rb)\right \}= \frac{\hbar\ww}{2} .
\end{split}
\end{equation}
Working in the quasistatic limit, the first integral above can be equated to the second by expressing the field as $\Eb=-\nabla\Phi$ and integrating by parts, i.e.,
\begin{equation}
    \int d^3\rb \Eb_\ww^*(\rb)\cdot\Eb_\ww(\rb) 
  =  \frac{4\pi}{\epsr}\int d^3\rb \Phi_\ww^*(\rb) \rho_\ww(\rb) ,
\end{equation}
where Gauss' law $\nabla\cdot\Eb=4\pi\rho/\epsr$ is invoked; for the 2D induced charge $\rho(\rb)=\rho(\Rb)\delta(z)$, we integrate over $z$ to write
\begin{equation}
    \int d^3\rb \Phi_\ww^*(\rb)\rho_\ww(\rb) 
    = -\frac{\ii\sigma_\omega^{(1)}}{\ww}\int d^2\Rb f_\Rb \Eb_\ww^*(\Rb)\cdot\Eb_\ww(\Rb) ,
\end{equation}
where the continuity equation $\rho = -(\ii/\ww)\nabla_\Rb\cdot\jb = -(\ii\sigma_\omega^{(1)}/\ww)\nabla_\Rb\cdot\ccpar{f_\Rb\Eb}$ was used to eliminate the charge density before again integrating by parts.
With the above result, the normalization condition becomes
\begin{equation}
\begin{split}
   & \Ree\clpar{\ccpar{\epsr + \frac{\ww}{2}\pd{\epsr}{\ww}} \frac{\sigma_\omega^{(1)}}{\ii\epsr\ww} + \frac{\ii}{2\ww}\pd{}{\ww}\ccpar{\ww\sigma_\omega^{(1)}} } \\
   & \times\int d^2\Rb f_\Rb \Eb_\ww^*(\Rb)\cdot\Eb_\ww(\Rb) = \frac{\hbar\ww}{2} ,
\end{split}
\end{equation}
and reduces to
\begin{equation}
    \Ree\clpar{ \frac{\sigma_\omega^{(1)}}{\ii\ww} + \frac{\ii}{2\ww}\pd{}{\ww}\ccpar{\ww\sigma_\omega^{(1)}} } \int d^2\Rb f_\Rb \Eb_\ww^*(\Rb)\cdot\Eb_\ww(\Rb) = \frac{\hbar\ww}{2}
\end{equation}
for a dispersionless environment.

The field in the ribbon can be expressed as $\Eb_\ww(\Rb)=\sum_n E_{n,k} \ub_{n,k}(x)\ee^{\ii k y}$, where $E_{n,k}$ is its amplitude and $\ub_{n,k} = \xx \partial_x \phi_{n,k}(x) + \yy \ii k \phi_{n,k}(x)$ are obtained by decomposing the potential in eigenmodes satisfying Eq.\ \eqref{eq:phi_eig}; the normalization condition then becomes, for a single mode, 
\begin{equation}
    \frac{1}{2\ww_{n,k}}\Imm\clpar{\left[ 2\sigma_\omega^{(1)} - \pd{(\ww\sigma_\omega^{(1)})}{\ww} \right]_{\ww=\ww_{n,k}}} L E_{n,k}^2 \xi_{n,k} = \frac{1}{2}\hbar\ww_{n,k}
\end{equation}
where $L$ is the mode length in $y$ and
\begin{equation}
    \xi^{(1)}_{n,k} = \int dx \abs{\ub_{n,k}(x)}^2 
\end{equation}
is the quantity depending on the mode function integral appearing in the main text, which scales as the ribbon width, i.e. $\xi^{(1)}_{n,k}\sim W$. From the above expressions, we isolate the field amplitude
\begin{equation}
    E_{n,k}^2 = \frac{\hbar\ww_{n,k}^2}{L\xi_{n,k} \Imm\clpar{\left[\sigma_\omega^{(1)}-\ww\partial_\ww\sigma_\omega^{(1)}\right]_{\ww=\ww_{n,k}}}} .
\end{equation}

\section{Plasmon absorption rates}

\subsection{Single-plasmon absorption}

\begin{figure}[!t]
\includegraphics[width=0.85\columnwidth]{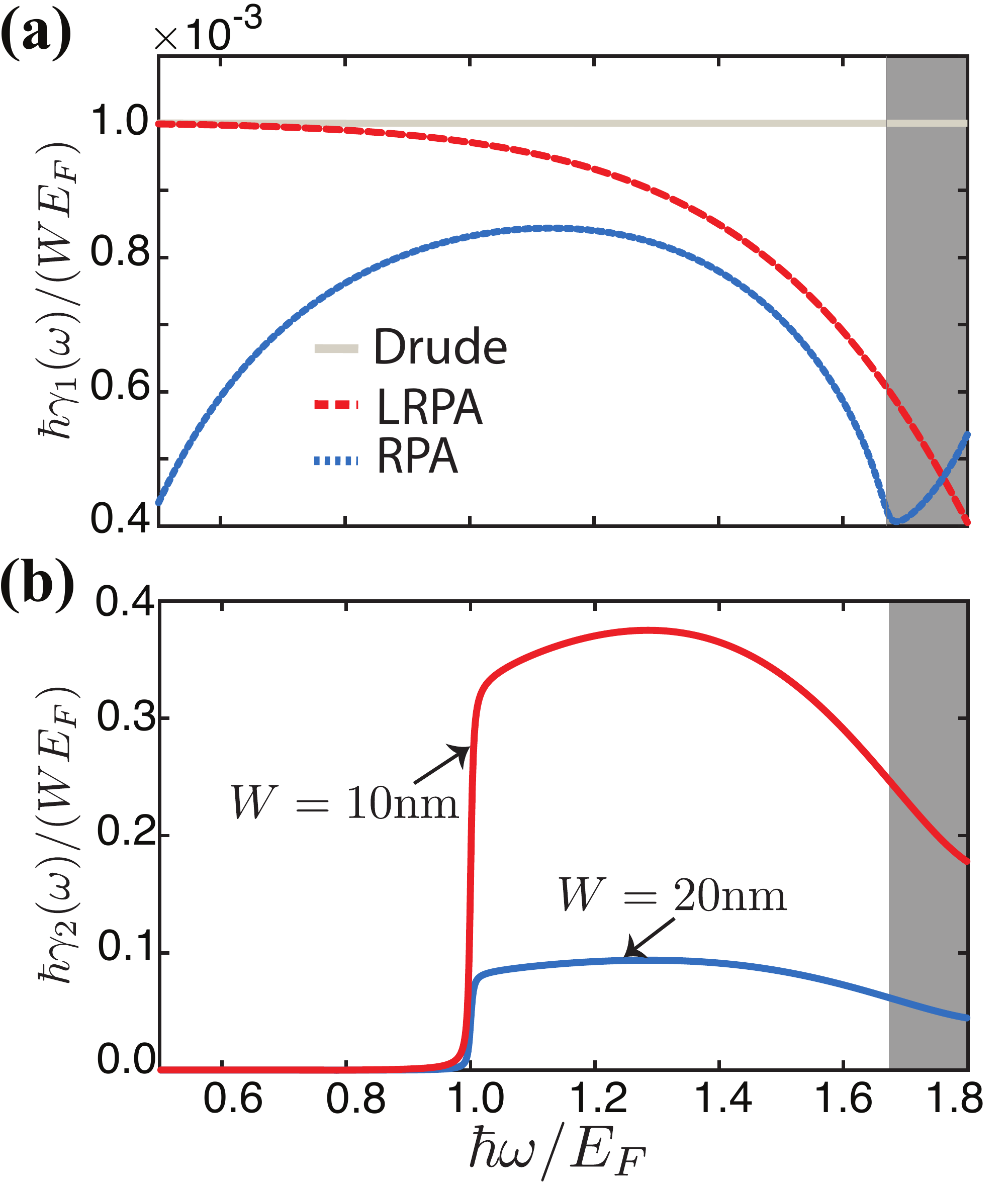}%
\caption{(a) Single-plasmon absorption rate vs. frequency calculated for the Drude model (grey line), LRPA (red line) and RPA with $kW=1$ (blue dots). The nanoribbon width was set  to $W=20$nm. (b) Normalized two plasmon absorption rate vs frequency calculated within LRPA for two different values of the nanoribbon width. In both plots we set $\EF=0.1$ and $\gamD=\EF/100$. The grey area indicates the frequencies above $\hbar\omega_{\rm plasma}\simeq 5\EF/3$ where plasmons cease to exist.}
\label{fig_loss_1plasmon}
\end{figure}

In order to estimate the single-plasmon decay rate we first evaluate the power absorbed by the ribbon, which is obtained from
\begin{equation} \label{eq:P_jE}
   P^{(1)} = \av{\int d^2\Rb \jb^{(1)}(\Rb,t)\cdot\Eb(\Rb,t)},
\end{equation}
where $\jb^{(1)}(\Rb,t)$ is the current to linear order and $\av{\dots}$ denotes the time average. For harmonic fields we insert $\jb^{(1)}(t)=\jb^{(1)}_\ww\ee^{-\ii\ww t} + {\rm c.c.}$ and $\Eb(t)=\Eb_\ww\ee^{-\ii\ww t} + {\rm c.c.}$
into Eq.\ \eqref{eq:P_jE} to find, after dropping the fast-oscillating terms which average to zero,
\begin{equation}
    P = 2 \Ree\clpar{\sigma^{(1)}_\ww} \int d^2\Rb \abs{\Eb_\ww(\Rb)}^2 ,
\end{equation}
having expressed the current in terms of the linear conductivity via Ohm's law $\jb^{(1)}_\ww=\sigma^{(1)}_\ww\Eb_\ww$. 
Equating the absorbed power with that dissipated at a rate $\gamma_1$ by the waveguide, we obtain the expression
\begin{equation}
    \gamma_1 = \frac{2\Ree\clpar{\sigma^{(1)}_\ww}}{\hbar \omega} \int d^2\Rb \abs{\Eb_\ww(\Rb)}^2=\frac{2\Ree\clpar{\sigma^{(1)}_\ww}}{\Imm\clpar{ \sigma_\ww^{(1)}-\ww\partial_\ww\sigma_\ww^{(1)}}},
\end{equation}
by using the mode normalization previously derived. The above equation establishes the rate of single-plasmon absorption in the ribbon; while for the Drude model the absorption rate coincides with the inelastic scattering rate, $\gamma_1(\omega)=\gamD$, 
the absorption rate predicted from the LRPA and RPA conductivities exhibit deviations as shown in Fig.\ref{fig_loss_1plasmon}(a), which arise from the different mode normalization factors. Importantly, even with the inclusion of the interband transition term, the dissipation rate does not exhibit a strong intrinsic frequency dependence, and remains on the same scale as the Drude rate. For this reason, we operationally define the total single-plasmon dissipation rate as a quantity set overall by the quality factor $Q$ according to $\gamma_1(\omega)=\omega/Q$, implicitly capturing all forms of dissipation. 

\subsection{Two-plasmon absorption}

To evaluate the two-plasmon absorption rate appearing in the model Hamiltonian, we compute the work done by the nonlinear current $\jb^{(3)}(\rb,t)$ on the plasmon field $\Eb(\rb,t)$ within a graphene layer occupying the $\Rb=(x,y)$ plane, such that the absorbed power reads  
\begin{equation}
    P^{(3)} = \av{ \int d^2\Rb \jb^{(3)}(\Rb,t)\cdot\Eb(\Rb,t) } .
\end{equation}
Decomposing the nonlinear current in its frequency components as $\jb^{(3)}(\Rb,t)=\jb^{(3)}_\ww\ee^{-\ii\ww t} + {\rm c.c.}$, and analogously the plasmon field, we obtain
\begin{equation}
    P^{(3)} = 2\Ree\clpar{\sigma^{(3)}_\ww}\int d^2\Rb \abs{\Eb_\ww(\Rb)}^4
\end{equation}
by writing $\jb^{(3)}_\ww=\sigma^{(3)}_\ww\Eb_\ww(\Rb)\abs{\Eb_\ww(\Rb)}^2$. The power absorption for a specific mode with index $n$ and wave vector $k$ is then given by
\begin{equation}
    P^{(3)}_{n,k} = 2\Ree\clpar{\sigma^{(3)}_{\ww_{n,k}}}E_{n,k}^4 L\xi^{(3)}_{n,k},
\end{equation}
where $\xi^{(3)}_{n,k} = \int dx \abs{\ub_{n,k}(x)}^4$. We associate the power absorption above with two-plasmon absorption at the rate $\Gamma^{(3)}(\ww_{n,k})$ according to $P^{(3)}=2\hbar\omega_{n,k}\Gamma^{(3)}$, which becomes
\begin{equation}
    \Gamma^{(3)} = 
    \frac{\hbar\omega^3_{n,k}}{ L }\frac{\Ree\clpar{\sigma^{(3)}(\ww_{n,k})}\xi^{(3)}_{n,k}}{\left[\Imm\clpar{\left[ \sigma_\ww^{(1)}-\ww\partial_\ww\sigma_\ww^{(1)} \right]_{\ww=\ww_{n,k}}}\xi^{(1)}_{n,k}\right]^2} .
\end{equation}
With the above result quantifying the two-plasmon absorption rate associated with a single mode, we finally obtain the rate of two-plasmon absorption per length as $\gamma_2=\Gamma^{(3)}L$, which is the TPA rate used in the effective model. The evaluated two-plasmon absorption rate normalized respect to the ribbon wavelength and the Fermi energy as function of the frequency is shown in Fig.\ \ref{fig_loss_1plasmon}(b) for two different ribbon widths, and is found to exhibit a large enhancement in the energy range between $\EF<\hbar \omega<\hbar\omega_{\rm plasma}$ that is considered in the main text to achieve a strong plasmon interaction.

\section{Plasmon scattering}\label{App.scattering}

\subsection{Effective model}
The effective model presented in the main text that describes plasmon-plasmon interactions is based on an expansion of the plasmon dispersion obtained in \ref{Sec.plasmon dispersion} around a given plasmon resonance $\omega_{\rm p}\simeq\omega_{n,k_{\rm p}}$ at momenta $k=\pm k_{\rm p}$. For plasmon frequencies away from the mode cutoff, we can separate the corresponding free plasmon momentum space Hamiltonian into right ($R$) and left ($L$) branches according to
\begin{equation}\label{Eq:breaking_int}
\begin{split}
&\mathcal{H}_0=\int dk\omega_k \hat a_k^{\dagger}\hat a_k\sim\\ & \int_{k_p-\Delta k}^{k_p+\Delta k} dk_R\omega_{k_R}\hat a_{k_R}^{\dagger}\hat a_{k_R}+\int_{-k_p-\Delta k}^{-k_p+\Delta k} dk_L\omega_{k_L}\hat a_{k_L}^{\dagger}\hat a_{k_L},
\end{split}
\end{equation}
where $ \omega_{k_\nu}=\bar\omega_{\rm p}\pm\bar v_{\rm g}k_\nu+\frac{\hbar}{2m}k_\nu^2$ is the approximate dispersion for the two branches labeled by $\nu=R,L$ ($\pm$), while we have defined $\bar\omega_{\rm p}=-v_{\rm g}k_{\rm p}+(\hbar/2m)k_{\rm p}^2$ and $\bar v_{\rm g}=v_{\rm g}-(\hbar/m)k_{\rm p}$.
To proceed further we transform the above Hamiltonian to position space by defining the Fourier transform $\hat a_{k_\nu }=\frac{1}{\sqrt{2\pi}}\int dy \hat a_\nu(y)e^{-ik_\nu y}$ in terms of the left and right bosonic operators $\hat a_\nu(y)$, which fulfill bosonic commutation rules $[\hat a_\nu(y),\hat a^{\dagger}_{\nu'}(y')]=\delta(y-y')\delta_{\nu\nu'}$ if the two branches are well-distinguished such that we can extend the limit of integration in k-space toward $\pm\infty$. Such a transformation leads to the free plasmon position space Hamiltonian presented in the main text,
\begin{equation}\label{Hwab}
    \mathcal{H}_0=-\sum_\nu\int_0^L d y \hat a_{\nu}^{\dagger}(y)\left(\frac{\hbar^2}{2m}\frac{\partial^2}{\partial y^2}\pm \ii\hbar\bar v_{\rm g}\frac{\partial}{\partial y}\right)\hat a_{\nu}(y) ,
\end{equation}
where we have omitted the frequency shift associated with the plasmon frequency term $\bar\omega_{\rm p}$. The plasmon nonlinearity is described as a local dissipative two-body interaction, which in momentum space reads as
\begin{equation}
    \mathcal{H}_{\rm I}=-\ii\frac{\gamma_2}{2}\int dq \int dk \int dp \,\hat a^{\dagger}(k+q)\hat a^{\dagger}(p-q)\hat a(p)\hat a(k) ,
\end{equation}
where $k$ and $p$ are the incoming momenta of the two plasmons and $q$ is the exchanged momentum. Proceeding as before, we separate the left and right contributions according to
\begin{equation}
    \mathcal{H}_{\rm I}=-\ii\frac{\gamma_2}{2}\sum_{\nu\nu'}\int dq dk_\nu dp_{\nu'} \hat a^{\dagger}(k_\nu+q)\hat a^{\dagger}(p_{\nu'}-q)\hat a(p_{\nu'})\hat a(k_\nu) ,
\end{equation}
where the integration limits for $k_\nu$ and $p_\nu$ are the same as in Eq.\ \eqref{Eq:breaking_int}, while $q$ takes values within the range $q\in[-\Delta q,\Delta q]$. Transforming the above Hamiltonian to position space, we finally obtain the interaction Hamiltonian presented in the main text,
\begin{equation}
    \mathcal{H}_{\rm I}=-\ii\frac{\gamma_2}{2}\sum_{\nu\nu'}\int dy \hat a_{\nu}^{\dagger}(y)\hat a_{\nu'}^{\dagger}(y)\hat a_{\nu}(y)\hat a_{\nu'}(y) .
\end{equation}

\subsection{Transmission and reflection coefficients}

The two-plasmon dynamics can be fully reconstructed by finding the eigenstates of the stationary Schr\"odinger equation $\mathcal{H}|\psi\rangle=\hbar\omega|\psi\rangle$, which can be solved using the generic ansatz 
\begin{equation}
|\psi\rangle=\sum_{\nu\nu'}\int dy_1 \int dy_2\psi_{\nu\nu'}(y_1,y_2)\hat a_{\nu}^{\dagger}(y_1)\hat a_{\nu'}^{\dagger}(y_2)
\end{equation}
taking into account all the field components. In our problem we are interested in the specific scenario of two counter-propagating plasmons, for which the equations describing the co-propagating component $\psi_{\nu\nu}$ are completely decoupled from their counter-propagating counterparts $\psi_{\nu\nu'}$. Considering that $\psi_{\rm RL}(y_1,y_2)=\psi_{\rm LR}(y_1,y_2)$, we define $\psi_{\rm RL}=\phi_{\rm R}$ and $\psi_{\rm LR}=\phi_{\rm L}$ to obtain a single wave equation depending on only one variable, i.e., the two-plasmon relative coordinate $\rho=|y_1-y_2|$, which explicitly reads 
\begin{equation}
\begin{split}
&\omega[\psi_{\rm R}(\rho)+\psi_{\rm L}(\rho)]=-\frac{\hbar}{m}\frac{\partial^2}{\partial \rho^2}[\psi_{\rm R}(\rho)+\psi_{\rm L}(\rho)]\\ &-2\ii\bar v_{\rm g}\frac{\partial}{\partial \rho}[\psi_{\rm R}(\rho)-\psi_{\rm L}(\rho)]-\ii\gamma_2\delta(\rho)[\psi_{\rm R}(\rho)+\psi_{\rm L}(\rho)].
\end{split}
\end{equation}\\

To solve the above equation, we consider a plasmon impinging from the left and make use of the ansatz 
\begin{equation}
\begin{split}
    \psi_{\rm R}(\rho)=&e^{ik\rho}\theta(-\rho)+te^{ik\rho}\theta(\rho)\\
    \psi_{\rm L}(\rho)=&te^{-ik\rho}\theta(-\rho) ,
\end{split}
\end{equation}
where $t=1+r$ ensures continuity of the solution at $r=0$ and $\theta(\rho)$ is the Heaviside step function; by imposing the boundary conditions coming from the delta function, we obtain transmission and reflection coefficients
\begin{equation}
    t=\left(1+\frac{\gamma_2}{4\bar v_{\rm g}+2\hbar k/m}\right)^{-1}\;\;\;\;\;\;\;\;r=-\left(1+\frac{4\bar v_{\rm g}+2\hbar k/m}{\gamma_2}\right)^{-1}.
\end{equation}
These coefficients can be rewritten in the same form as the main text, $r=-\left[1+2\pi\lambda_{\rm a}/\lambda_{\rm p}\right]^{-1}$ and $t=\left[1+\lambda_{\rm p}/(2\pi\lambda_{\rm a})\right]^{-1}$ by defining the absorption length
\begin{equation}
    \lambda_{\rm a}=\frac{4\bar v_{\rm g}}{k_{\rm p}\gamma_2}+\frac{2\hbar}{m\gamma_2}=\frac{4 v_{\rm g}}{k_{\rm p}\gamma_2}-\frac{2\hbar}{m\gamma_2} ,
\end{equation}
where we have expressed the RHS in terms of the original group velocity at the plasmon resonance.

\bibliography{refs}

\end{document}